\documentclass[10pt, twocolumn,
							 superscriptaddress,
							 english,
							 prx,
							 showpacs,
							 floatfix,
							 aps
							]{revtex4-2}
\usepackage{graphicx} 
\usepackage[utf8]{inputenc}
\usepackage{amsmath}
\usepackage{amssymb}
\usepackage{graphicx}
\usepackage{xspace}
\usepackage{xfrac}
\usepackage{xcolor}
\usepackage[backref=none,bookmarksnumbered=true,bookmarks=true,bookmarksopen=true,colorlinks=true,
citecolor=blue,linkcolor=blue,anchorcolor=green,urlcolor=blue,unicode=false]{hyperref}
\usepackage{ulem}
\usepackage{siunitx}
\usepackage[separate-uncertainty = true,multi-part-units=single]{siunitx}
\usepackage{romannum}
\usepackage[overload]{textcase} 

\newcommand{\NbS}{$2H$-NbS$_2$}
\newcommand{\NbSe}{$2H$-NbSe$_2$}
\newcommand{\nbse}{NbSe$_2$}

\newcommand{\Vbias}{$V_{\text{bias}}$}
\newcommand{\It}{$I_{\text{t}}$}
\newcommand{\dIdV}{d$I$/d$V$}
\newcommand{\didv}{\ensuremath{\mathrm{d}I/\mathrm{d}V}\xspace}

\newcommand{\beginsupplement}{%
	\setcounter{table}{0}
	\renewcommand{\thetable}{S\arabic{table}}%
	\setcounter{figure}{0}
	\renewcommand{\thefigure}{S\arabic{figure}}%
	\setcounter{equation}{0}
	\renewcommand{\theequation}{S\arabic{equation}}%
	\setcounter{section}{0}
	\renewcommand{\thesection}{S\arabic{section}}%
}

\date{\today}

\begin{document}

\title{Probing crystal-field modulations with magnetic adatoms on the incipient charge-density-wave superconductor \NbS \\}

\author{Werner M. J. van Weerdenburg}
 \email{werner.van.weerdenburg@fu-berlin.de}
\affiliation{Fachbereich Physik and Halle-Berlin-Regensburg Cluster of Excellence (CCE), Freie Universit\"at Berlin, Arnimallee 14, 14195 Berlin, Germany.}

\author{Margarete Huisinga}
\affiliation{Fachbereich Physik and Halle-Berlin-Regensburg Cluster of Excellence (CCE), Freie Universit\"at Berlin, Arnimallee 14, 14195 Berlin, Germany.}

\author{Constantin Flommersfeld}
\affiliation{Fachbereich Physik and Halle-Berlin-Regensburg Cluster of Excellence (CCE), Freie Universit\"at Berlin, Arnimallee 14, 14195 Berlin, Germany.}

\author{Lisa M. R\"{u}tten}
\affiliation{Fachbereich Physik and Halle-Berlin-Regensburg Cluster of Excellence (CCE), Freie Universit\"at Berlin, Arnimallee 14, 14195 Berlin, Germany.}

\author{Katharina J. Franke}
\affiliation{Fachbereich Physik and Halle-Berlin-Regensburg Cluster of Excellence (CCE), Freie Universit\"at Berlin, Arnimallee 14, 14195 Berlin, Germany.}

\begin{abstract}
The interplay between multiple quantum phases in layered materials may lead to incipient quantum behavior, where the material's ground state is close to a phase transition and sensitive to local disorder. The transition metal dichalcogenide material \NbS\ exhibits incipient charge-density-wave behavior along with a well-developed superconducting state, creating a scenario where the local lattice instabilities play a crucial role. Here we present how an individual magnetic atom on \NbS\ can be applied as a local sensor to reveal hidden crystal-field modulations. By manipulating the adatom across the surface with the tip of a scanning tunneling microscope, we measure variations in the Yu-Shiba-Rusinov (YSR) excitation spectra and map the local environment around an intrinsic point defect. We find that while the superconducting state is spatially uniform, the YSR excitation energy strongly depends on the position of the atom. We determine that the main contribution to this effect originates from variations in the local crystal-field environment. These results establish a new approach to investigate crystal-field modulations at the atomic scale and reveal how defects and lattice instabilities shape the atomic landscape of an incipient charge-density-wave material.
\end{abstract}

\maketitle

\section{Introduction}
The interplay of correlated quantum phases is at the forefront of condensed matter research, with a strong focus on layered materials where interactions often induce multiple coexisting phases at low temperatures. Of particular interest is the interplay of superconductivity and charge-density ordering, where it is debated whether the two phenomena compete or reinforce one another \cite{Chang2012, Wang2023, Suderow2005, Kiss2007, Borisenko2009, Cho2018}. In several superconducting materials, the emergence of a charge-density wave (CDW) has been reported to be accompanied by a pair-density wave, underscoring the interaction between multiple phases \cite{Liu2021, Aishwarya2023, Gu2023, Liu2023}. In real materials that host sources of disorder, such as strain or defects, the balance of interactions may locally be disturbed with substantial consequences for the ground state of the system \cite{Sokolovic2019, Guo2024, Xiang2025, Ge2026}. Probing the influence of atomic-scale defects offers a unique opportunity to study the interplay of correlated phases. As the defects locally modify the electronic potential, they can alter the relative stability of two phases, providing insight into whether they compete, coexist or even cooperate. 

Transition metal dichalcogenides (TMDC) are particularly well suited to study coexisting orders. Among them is \NbSe, where superconductivity coexists with an incommensurate CDW ($T_{\text{CDW}}$ = \SI{33}{K}, $T_{\text{c}}$ = \SI{7.2}{K}) \cite{Suderow2005, Kiss2007, Borisenko2009, Cho2018}. The bulk phase of \NbSe\ has been investigated extensively to determine the origin of the CDW phase \cite{Rossnagel2001, Johannes2006, Arguello2015}, how its multiband character affects superconductivity \cite{Yokoya2001, Rodrigo2004, Noat2010, Noat2015, Sanna2022} and how atomic-scale defects interact with the CDW phase \cite{Oh2020, Sheng2024, Rutten2025}. Recent studies utilizing atomic-scale Josephson spectroscopy also report a pair-density-wave phase in \NbSe\ \cite{Liu2021, Cao2024}, suggesting a symbiotic relation between the low-temperature phases \cite{Machida1981}. In contrast, the isostructural analog \NbS\ stands out among the TMDC materials, as it hosts superconductivity below $T_c \approx$ \SI{6}{K} without the presence of a long-range CDW phase \cite{Fisher1980, Leroux2012, Guillamon2008a}, despite its structural, electronic and phononic similarities to \NbSe\ \cite{Suzuki2005, Majumdar2020}. This contrasting behavior has been attributed to subtle differences in the electron-phonon interactions, where anharmonic phonon effects place the system on the verge of a CDW phase transition \cite{Leroux2012, Heil2017, Bianco2019}. This raises the question whether defects in an incipient CDW material introduce lattice instabilities and/or influence the balance between the correlated phases \cite{Wen2020}.

To address this question at the atomic scale, the interaction between superconductivity and magnetic adatoms can provide insight. The exchange coupling between the unpaired spins of a magnetic impurity and the superconductor induces sub-gap excitations and gives rise to sharp resonances inside the superconducting gap known as Yu-Shiba-Rusinov (YSR) states \cite{Yu1965,Shiba1968, Rusinov1968}. The excitation energy has been shown to be highly sensitive to small changes in the exchange coupling strength \cite{Franke2011, Hatter2015, Cornils2017, Farinacci2018, Malavolti2018, Huang2020} and the local density of states \cite{Liebhaber2020}. Changes in the superconducting order parameter or potential scattering are also predicted to affect the YSR energy \cite{Kiendl2017, Babkin2022}. As a result, the local interactions between magnetic adatoms and substrate may be used as sensor to reveal otherwise hidden structural and electronic instabilities. 

Here, we utilize individual Fe atoms on \NbS\ as sensors to reveal defect-induced lattice distortions in the crystal. By manipulating the Fe atoms across the surface with the tip of a scanning tunneling microscope (STM), we map the nanometer-scale environment of an intrinsic point defect and track the position-dependent changes of the YSR excitation energy. Despite the spatially homogeneous superconducting state, we find a surprisingly large YSR energy variation that corresponds to $\sim$ 70\% of the inner gap size. By quantifying the effect of density-of-states variations and potential scattering, we show that the main origin of this atomic-scale heterogeneity stems from variations in the crystal-field environment of the adatom. These lattice distortions indicate that intrinsic defects in an incipient CDW material can strongly modify its atomic-scale structural properties. 

\section{Results}
Atomically-flat, pristine surfaces of \NbS\ are obtained by cleaving a bulk crystal \textit{in-situ} and transferring the crystal directly into the STM. Topographic images of the surface (Fig.\,\ref{fig:Fig1}a) reveal the triangular atomic lattice of the top sulfur (S) layer, as well as several intrinsic point defects. We identify four common defect types (\Romannum{1} - \Romannum{4}), as shown in Fig.\,\ref{fig:Fig1}b. Based on their central position with respect to the lattice of S atoms, we assign the defect type \Romannum{1} and \Romannum{2} to individual S vacancies in the top and bottom row of the \NbS\ lattice (see Supplemental Material (SM) and Fig.\,S1). This assignment is corroborated by defects in other TMDC materials such as \nbse, where chalcogen vacancies are abundant and have a similar appearance \cite{Oh2020}. The defect types \Romannum{3} and \Romannum{4} are less abundant and could originate from interstitial defects or a vacancy at the niobium (Nb) site. Lastly, fainter defect types, indicated with arrows in Fig.\,\ref{fig:Fig1}a, can be identified and likely correspond to sub-surface S vacancies in the \NbS\ layers below the exposed surface layer. 

Additionally, standing wave patterns are clearly visible in the vicinity of the defects in Fig.\,\ref{fig:Fig1}a. From spatially-resolved \dIdV\ conductance measurements at different bias voltages, we find that their periodicity depends on energy (see Fig.\,\ref{fig:SIFig_QPImap}), indicating that these modulations should be interpreted as Friedel oscillations due to defect-induced electronic scattering rather than local charge-density-wave modulations \cite{Wen2020}. However, local patterns may also occur as a result of strain in the material, which is common for cleaved layered crystals and can have important implications for the CDW formation in TMDC materials \cite{Soumyanarayanan2013, Gao2018, Zhang2025}.

\subsection{Superconductivity and intrinsic defects of \NbS\ }
\begin{figure*}\centering	\includegraphics{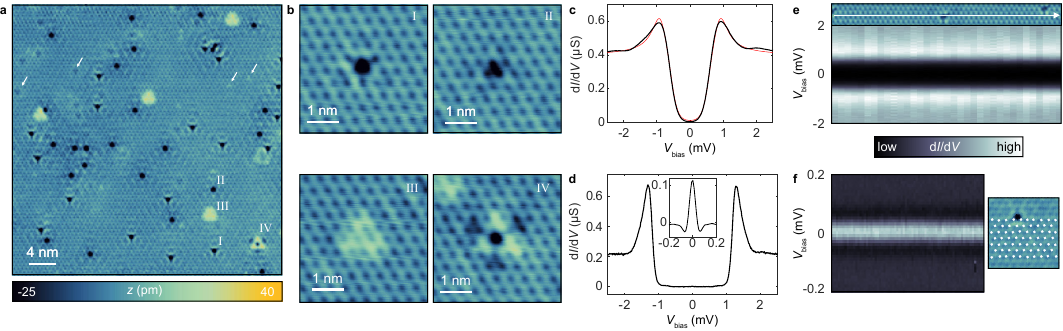}
	\caption{Superconductivity and intrinsic defects of \NbS. a) Topographic STM image of \NbS\ with several types of intrinsic defects, labeled with \Romannum{1} - \Romannum{4} and white arrows (\It = \SI{500}{\pA}, \Vbias = \SI{50}{\mV}). b) Close-up STM images of four defect types (\It\ = \SI{1}{\nA}, \Vbias\ = \SI{100}{\mV}). c) Tunneling spectrum showing the superconducting gap of \NbS (black), spatially averaged over an area of \qtyproduct{20 x 20}{\nm} and fitted with a two-gap Dynes function (red line). d) Tunneling spectrum recorded with a superconducting tip on \NbS, spatially averaged over 10 sites. Inset: Josephson conductance peak recorded for a V-NbS$_2$ junction at a normal-state conductance of $G_N$ = \SI{1.6}{\micro S} and averaged over all sites indicated in f). e) \dIdV\ spectroscopy along a \SI{20}{nm} line (top image), revealing a spatially-homogeneous gap in the presence of defects. f) Spatially-resolved Josephson conductance at $G_N$ = \SI{1.6}{\micro S} on \NbS\ (see small topographic image for positions). The spectra were stabilized at \Vbias\ = \SI{5}{\mV}, (c-e) \It\ = \SI{2}{nA}, $V_{\text{mod}}$ = \SI{50}{\micro V}) or (d) \It\ = \SI{1}{nA} and (d inset, f) \It\ = \SI{8}{nA}, and recorded with (c,e) a Au-coated tip or (d,f) a V-coated tip.} 
	\label{fig:Fig1}
\end{figure*}

Next, we probe the superconducting state of \NbS\ and how it is influenced by the various defects. Spectroscopic measurements at $T$ = \SI{1.1}{\K} with a gold-coated tungsten tip show a pronounced superconducting gap with broad coherence peaks (Fig.\,\ref{fig:Fig1}c). The lineshape can be fitted with a Dynes-broadened BCS gap function \cite{Dynes1978} that includes two gaps: $\Delta_1 =$ \SI{0.60(0.04)}{meV}, $\Delta_2 = $ \SI{0.84(0.03)}{meV}, consistent with the multi-band nature of superconductivity in \NbS\ \cite{Guillamon2008a, Sanna2022}. To enhance the spectroscopic energy resolution, we coat the tip with vanadium (V), which exhibits a gap of $\Delta \approx $ \SI{0.75}{meV}. All spectral features of the substrate are then shifted by the tip gap, leading to a total observed gap of $\Delta_{\text{max}} = \Delta_{\text{NbS}_{2}} + \Delta_{V} \approx $ \SI{1.35}{\meV}, as shown in Fig.\,\ref{fig:Fig1}d. Spatially resolved spectroscopic measurements along a line in Fig.\,\ref{fig:Fig1}e demonstrate that the superconducting gap remains constant and shows no significant changes or in-gap conductance in the presence of an individual defect. This suggests that the intrinsic defects of this material are non-magnetic (also see Fig.\,\ref{fig:S1_Defects}). We note that small modulations can be observed in the height of the coherence peaks, which may appear due to variations in the tunneling matrix elements in a multi-band superconductor \cite{RubyPb15, Hanaguri2026}. 

Another asset of a superconducting tip is that it can be used to form an atomic-scale Josephson junction by bringing the tip sufficiently close to the sample. A sharp conductance peak at zero bias then indicates Cooper pair tunneling (see inset in Fig.\,\ref{fig:Fig1}d). As magnetic impurities lead to a suppression of the Josephson peak \cite{Randeria2016, Trahms2023}, its height can be used as a complementary sensor for the perturbation of the superconducting properties. Additionally, periodic modulations of the Josephson-peak height may reflect a pair-density wave, which often comes along with a charge-density wave in superconducting materials \cite{Hamidian2016, Liu2021}. As shown in Fig.\,\ref{fig:Fig1}f, the Josephson conductance probed around an individual defect shows no significant variations, indicating that the Cooper pair density and pairing strength of \NbS\ can be considered uniform. 

\subsection{Magnetic Fe adatoms as atomic-scale sensors}
\begin{figure} \centering	\includegraphics[width=\linewidth]{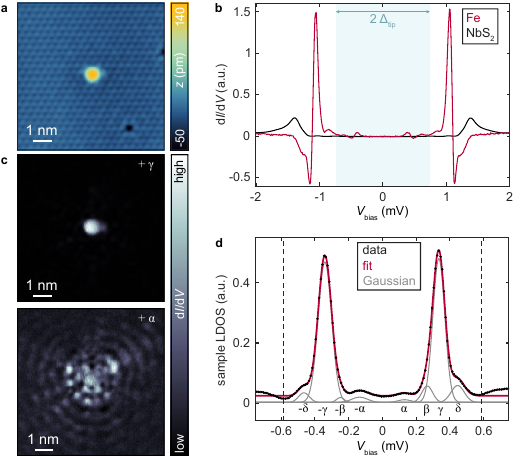}
\caption{Yu-Shiba-Rusinov states of an individual Fe atom on \NbS. a) Topographic STM image of an individual Fe atom on \NbS\ (\It = \SI{200}{\pA}, \Vbias = \SI{5}{\mV}). b) \dIdV\ spectrum of Fe on \NbS\ (red) and the substrate (black), measured with a superconducting V tip (stabilized at \Vbias\ = \SI{5}{\mV},  \It\ = \SI{2}{nA} (red) and \SI{500}{pA} (black)). The size of the superconducting gap of the tip is indicated with blue shading. c) \dIdV\ images of the $+\gamma$ (top) and $+\alpha$ (bottom) state, recorded in constant-contour mode at the height profile presented in (a) ($V_{\text{mod}}$ = \SI{30}{\micro V}, see SM for more details). d) Fe spectrum (black) after numerical deconvolution to remove the influence of tip DOS, fitted with multiple Gaussian profiles (red) to identify up to four pairs of YSR states $\alpha$, $\beta$, $\gamma$ and $\delta$ (grey). Dashed lines indicate the onset of the superconducting gap edge of \NbS.} 
	\label{fig:Fig2}
\end{figure}

Next, we deposit individual Fe atoms on the surface of \NbS, as shown in Fig.\,\ref{fig:Fig2}a, and characterize their YSR states before utilizing them as atomic sensors. We find that Fe atoms adsorb at the hollow sites of the top S layer and occupy two distinct adsorption sites, distinguished by the presence or absence of an underlying Nb atom. Given that the layer of Nb atoms is not visible in topographic images, we cannot experimentally assign which type of Fe atom is above a Nb atom. However, the two adsorption sites can be readily distinguished by their distinct YSR-state fingerprints (see Fig.\,\ref{fig:SIFig_Fe_bindingsites}). For this work, we focus on the Fe type presented in Fig.\,\ref{fig:Fig2}b, which is characterized by a pair of high-intensity YSR states flanked by strong negative differential conductance at the higher energy side. Additionally, there are less-pronounced pairs of YSR states at lower energies. Due to the superconducting nature of the tip, the YSR states are not only shifted by $\Delta_\mathrm{tip}$, but also lead to thermal replica at $|eV|= \Delta_\mathrm{tip}-\epsilon_\mathrm{YSR}$.
To remove the effect of the superconducting density-of-states (DOS) of the tip, we numerically deconvolve the spectra (see SM and Fig.\,\ref{fig:SIFig_Deconv} for details). After the deconvolution procedure (e.g., Fig.\,\ref{fig:Fig2}d), we identify up to four YSR states and label the peaks as $\pm$ $\alpha$, $\beta$, $\gamma$, and $\delta$. 

The four YSR states originate from the singly-occupied crystal-field split $d$ levels of the Fe atom \cite{Ruby2016}. The strongest intensity of the $\gamma$ state indicates the largest wavefunction overlap with the STM tip. Furthermore, the spatially-resolved YSR wavefunctions presented in the \didv maps in Fig.\,\ref{fig:Fig2}c, show that the $\gamma$ state is strongly localized, suggesting that it originates from the $d_{z^2}$ orbital. The other YSR states are less intense and exhibit longer-ranged oscillatory patterns (see Fig.\,\ref{fig:SIFig_YSRmaps} for details). The symmetry of the pattern appears complex and reflects the combination of the local crystal field and the shape of the Fermi surface \cite{Liebhaber2020}. 

\begin{figure*}\centering	\includegraphics{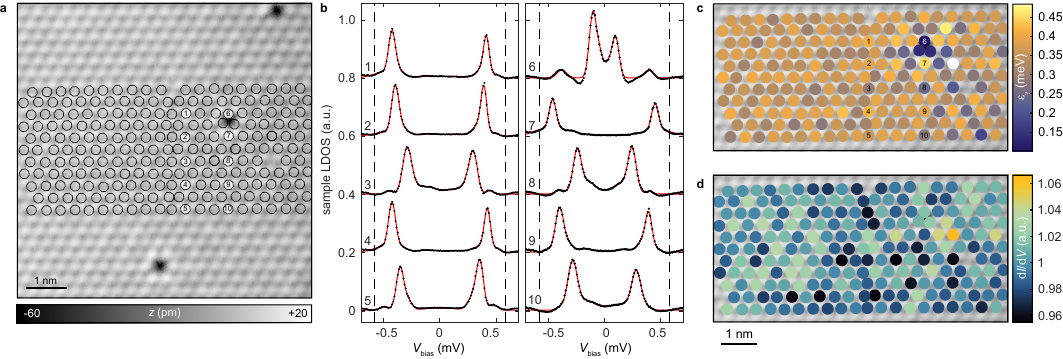}
    \caption{Spatial variation of Yu-Shiba-Rusinov energies on \NbS. a) Topographic STM image of bare \NbS\ indicating the 225 equivalent binding sites (black circles) where an individual Fe atom was placed and probed (\It = \SI{1}{\nA}, \Vbias = \SI{5}{\mV}). b) Selection of deconvolved \dIdV\ spectra recorded on Fe in 10 binding sites, as labeled in (a). Each spectrum was fitted with multiple Gaussian peaks (red) and vertically offset for clarity. Dashed lines indicate the edges of the superconducting gap of \NbS. c) Extracted YSR excitation energy of the $+\gamma$ state of the Fe sensor atom positioned at each probed binding site. d) Average \dIdV\ signal at each binding site, extracted from a constant-current \dIdV\ image recorded at \Vbias = \SI{5}{\mV} and \It\ = \SI{1}{nA} (see Section \ref{sec:dIdVanalysis}). The color scale represents the percentage of observed variation.} 
	\label{fig:Fig3}
\end{figure*}

While the YSR state spectrum presented in Fig.\,\ref{fig:Fig2}b is typical in terms of relative intensity distributions among the four states for many of the Fe atoms in this binding site, we find that the exact energetic position of the states is highly sensitive to the local environment of the atom. To unravel the origin for these variations, we utilize the tip of the STM to position the Fe atom in equivalent binding sites on a selected area (see SM for details). In turn, the highly sensitive YSR states may reveal inhomogeneities that are hidden in the usual imaging. With this approach, we investigated 225 equivalent hollow sites, as indicated by the black circles on the topographic image in Fig.\,\ref{fig:Fig3}a, and recorded and analyzed their YSR state spectra as detailed in Section \ref{sec:methods}. The precisely chosen adsorption sites for the Fe atom cover an area of approximately \SI{8}{nm} by \SI{4}{nm}, that includes only one clearly visible defect of type \Romannum{1}. The rest of the area does not reveal pronounced variations in the topographic image. 

A selection of deconvolved spectra recorded centrally on an individual Fe atom positioned in binding sites 1 – 10 (arbitrarily selected) is presented in Fig.\,\ref{fig:Fig3}b. Here, we focus on the energy interval within the superconducting gap (indicated by dashed lines as the coherence peaks are suppressed) to highlight the differences in the energy position of the YSR states. The exemplary spectra reveal strong variations of the YSR states, most obviously reflected in the shift of the strongest-intensity state ($\gamma$) through almost the entire gap range. Most notably, the Fe atom in position 6, positioned directly at the defect site, shows a strongly modified spectrum where the highest intensity peaks are close to the Fermi level, while other sites display this pair of peaks closer to the gap edges. While a strong influence may be expected from a different crystal-field environment at the defect site due to a missing S atom, binding sites further away from the defect, where the topographic image has no features, also show significant variations in the energetic position of the $\gamma$-state ($\epsilon_{\gamma}$). By extracting the position of the strongest YSR state $\epsilon_{\gamma}$ from the deconvolved data across all probed binding sites, we are able to map the variation in YSR energy as shown in Fig.\,\ref{fig:Fig3}c. The defect is clearly identifiable as the binding sites where $\epsilon_\gamma$ is found at the lowest energies. However, the map crucially reveals that there are significant variations in $\epsilon_\gamma$ even at binding sites several nanometers away from the defect. Across the map, $\epsilon_\gamma$ shifts between \SI{0.10}{meV} and \SI{0.49}{meV}, thereby spanning $\sim$ 70\% of the inner gap size of \NbS\ \cite{Guillamon2008a}. The variation is comparable in magnitude to that observed for Fe atoms on \nbse\, where the CDW is fully developed \cite{Liebhaber2020}. We further note that the tip position has no significant effect on the extracted $\epsilon_\gamma$ for a given binding site (see Fig.\,\ref{fig:SIFig_tipinfluence}). 

\subsection{Origin of spatial variations in \NbS}
To understand the physical origin of the observed variation, it is insightful to identify the several components that may influence the energy of a YSR excitation. The excitation energy for a classical spin is given by \cite{Yu1965, Shiba1968, Rusinov1968, Balatsky2006}
\begin{equation}
    \epsilon_\mathrm{YSR} = \Delta \frac{1-A^2 + B^2}{\sqrt{(1-A^2+B^2)^2 + 4A^2}}.
    \label{eq:YSRenergy}
\end{equation} 
Here, $A = \frac{1}{2}\pi S \rho_0 J $ represents the exchange coupling between the unpaired spin of the magnetic impurity and the superconductor, and $B = \pi \rho_0 K$ describes the potential scattering. Therefore, the YSR energy is sensitive to variations in the superconducting order parameter $\Delta$, the normal-state density of states at the Fermi level $\rho_0$, the potential scattering coefficient $K$ and the exchange coupling $J$. In the following, we will discuss each of these terms for the case of Fe atoms on \NbS\ and experimentally determine which term causes the large variation of $\epsilon_{\gamma}$. 

Based on the results presented in Fig.\,\ref{fig:Fig1}d and \ref{fig:Fig1}f, the superconducting order parameter $\Delta$ can be considered as constant, whereas the quasiparticle interference patterns in Fig.\,\ref{fig:Fig1}a indicate that the density of states is non-uniform across the surface. To ascertain the relation between the local density of states (LDOS) and the variation in $\epsilon_{\gamma}$, we mapped the \dIdV\ signal close to the Fermi level (\Vbias\ = \SI{5}{\mV}) in the investigated area presented in Fig.\,\ref{fig:Fig3}a. The corresponding Friedel oscillations in Fig.\,\ref{fig:SIFig_dIdVmap} are isolated from the topographic contribution at the atomic periodicity by means of Fourier-filtering (see Section \ref{sec:dIdVanalysis} for details). Lastly, the signal at each binding site is evaluated by averaging within a radius of 150 pm. The resulting discretized image is represented in Fig.\,\ref{fig:Fig3}d and shows that the LDOS varies by approximately $\pm$ 5\%, with stronger modulations in the vicinity of the defect. However, there is no apparent correlation between the patterns observed in Fig.\,\ref{fig:Fig3}c and \ref{fig:Fig3}d. Similarly, the \dIdV\ signal was analyzed at eight energies between \SI{-20}{\mV} and \SI{20}{\mV} to check for correlations in a larger energy window (see Fig.\,\ref{fig:SIFig_dIdVmap_disc}).

\begin{figure} \includegraphics[width=\linewidth]{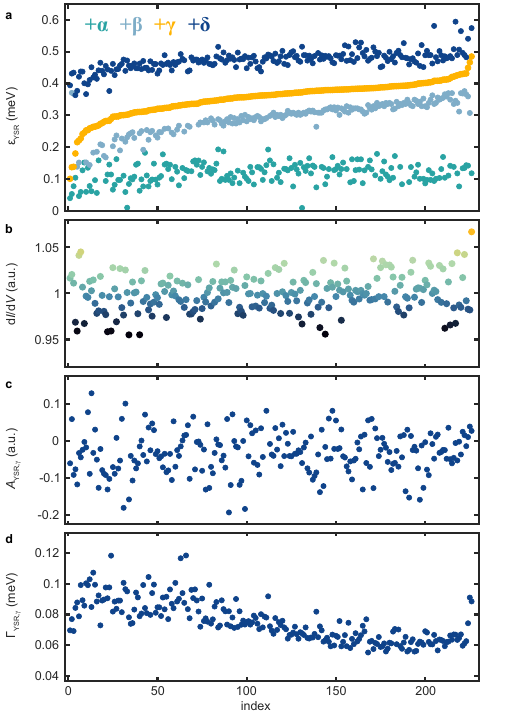}
    \caption{Yu-Shiba-Rusinov state variation. a) Scatter plot of the YSR state energies for $\alpha$, $\beta$, $\gamma$ and $\delta$ for each of the probed binding sites in Fig.\,\ref{fig:Fig3}. The binding site index is sorted by the energy of $+\gamma$. b) Corresponding scatter plot of the average \dIdV\ signal at each binding site, extracted from a constant-current \dIdV\ image recorded at \Vbias\ = \SI{5}{\mV}. c) Corresponding scatter plot of the YSR state asymmetry of $\gamma$, extracted from the intensities of $+\gamma$ and $-\gamma$. d) Corresponding scatter plot of YSR state width (FWHM) of $+\gamma$.} 
	\label{fig:Fig4}
\end{figure}

To further evaluate the relation between $\epsilon_\gamma$ and the parameters in equation \ref{eq:YSRenergy}, we sort the spectra according to their value of $\epsilon_\gamma$ and assign an index to each binding site. This yields a one-dimensional representation of the data in Fig.\,\ref{fig:Fig3}a where the increasing energy trend of $\epsilon_\gamma$ can be directly compared to other quantities. Figure \ref{fig:Fig4}a also shows the corresponding energies of the other YSR states ($\epsilon_\alpha$, $\epsilon_\beta$ and $\epsilon_\delta$), revealing a clear correlation between the $\epsilon_\gamma$ and $\epsilon_\beta$. Interestingly, the variation of $\epsilon_\gamma$ is much larger than the energy variation of the other states, possibly related to the orbital nature of the $\gamma$-state. We also note that not every spectrum in the data set contains four distinct pairs of YSR states, either due to state degeneracy or weak intensity, resulting in some indices with only three scatter points. Similarly, we plot the corresponding \dIdV\ signal from Fig.\,\ref{fig:Fig3}d at every binding site index in Fig.\,\ref{fig:Fig4}b and observe no significant slope in the LDOS variation with increasing $\epsilon_\gamma$. This rules out the scenario that the observed YSR variation can trivially be assigned to the influence of Friedel oscillations. 

The role of potential scattering is reflected in the asymmetry in YSR intensity at both polarities. Therefore, we define the YSR asymmetry as $A_{\text{YSR}} = (I(+\gamma) - I(-\gamma)) / (I(+\gamma) + I(-\gamma))$, where $I$ is the amplitude of the YSR state, and plot the resulting value for the $\gamma$ state in Fig.\,\ref{fig:Fig4}c. This shows that on average, all the YSR states are approximately symmetric, and again, there is no clear trend observed, suggesting that variations in potential scattering are not the main cause of the YSR energy variations. As a complementary measure, we look at the full-width half maximum of the $+\gamma$ YSR state ($\Gamma_{\text{YSR}}$, Fig.\,\ref{fig:Fig4}d) and find a slight downward trend with increasing $\epsilon_\gamma$, with values between \SI{60}{\micro V} and \SI{120}{\micro V}. This finding is interesting, since the width of a YSR state has been proposed as a measure for the potential scattering due to disorder \cite{Kiendl2017, Babkin2022} and could therefore indicate a small contribution due to potential scattering in a quasi two-dimensional superconductor. However, the observed variation in $\Gamma_{\text{YSR}}$ is small compared to the variation of $\epsilon_\gamma$. Alternatively, the effect can be ascribed to lifetime of the YSR state, since the relaxation rates of the excitation also depend on its energy \cite{Ruby2015}. Based on the YSR asymmetry and width, we can therefore confirm that potential scattering is not the main mechanism at play. 

Having analyzed the LDOS variations and potential scattering, we propose that the remaining term in equation \ref{eq:YSRenergy}, the exchange coupling $J$, is spatially varying. While a modified $J$ at the S vacancy site is readily understood due to the strongly perturbed atomic-scale environment, our analysis further indicates variations in regions, where neither the STM topography nor LDOS exhibit correlated changes. Since the exchange coupling is given by the wavefunction overlap between the adatom's atomic orbitals and surrounding crystal field, these observations indicate the presence of lattice reconstructions in \NbS\ that are invisible in the STM topography and LDOS. 

Based on this conclusion, we can infer the atomic-scale response of an incipient CDW material to intrinsic defects. In particular, we believe that small displacements of the Nb atoms may play a crucial role as they are expected to have the strongest impact on the $\gamma$ state derived from the $d_{z^2}$ orbitals. This interpretation is consistent with studies of the CDW phase in \NbSe, where the structural distortion is dominated by displacements of the Nb sublattice \cite{Malliakas2013, Lian2018}, as well as calculations predicting CDW structures in \NbS\ that likewise involve substantial Nb displacements and soft phonon modes including Nb vibrations \cite{Leroux2012, Bianco2019}.

The origin of the Nb displacements remains speculative. One possible scenario is that S vacancies induce a long-range relaxation of the lattice, which may be sensed by the Fe atoms even a few nm away from the defect site. The presence of multiple vacancies, however, would shape a complex energy landscape and prevent the formation of a clear wave pattern with a single periodicity, in contrast to a full-formed CDW system. Another possibility is that the lattice responds to less local perturbations such as defects in subsurface layers, stacking faults, polytypism or strain \cite{Leroux2018, Witteveen_2021}. However, we consider this explanation less likely as the LDOS in the investigated area does not indicate these features and the superconducting state also remains unchanged. 
 
\section{Conclusions}
In conclusion, we have demonstrated that individual Fe adatoms can serve as highly sensitive atomic-scale probes of the local crystal-field environment in the superconducting transition-metal dichalcogenide \NbS. By combining atom manipulation with high-resolution YSR spectroscopy, we mapped the spatial evolution of the exchange interaction across hundreds of equivalent adsorption sites surrounding an intrinsic defect. While conventional STM, including topography, \dIdV\ spectroscopy, and Josephson measurements, reveal a remarkably homogeneous surface and superconducting state, the YSR excitation energies exhibit pronounced variations over nanometer length scales. By systematically evaluating the possible contributions of superconducting gap variations, LDOS modulations, and potential scattering, we identify spatial variations of the exchange coupling as the dominant origin of the observed YSR-energy shifts.

Beyond establishing a new atomic-scale sensing approach, our results provide further insight into the microscopic properties of \NbS. Despite the absence of CDW order and the spatially homogeneous superconducting gap, the crystal hosts hidden structural inhomogeneities that strongly modify the local crystal field. The pronounced sensitivity of the $d_{z^2}$-derived YSR state suggests that these distortions primarily involve subtle displacements of the Nb atoms, consistent with the soft lattice dynamics that place \NbS\ in close proximity to a CDW instability. While our measurements do not reveal static CDW order, they indicate that intrinsic defects locally reshape the lattice over nanometer length scales. These defect-induced lattice relaxations act as an atomic-scale expression of the incipient CDW phase and highlight that the structural susceptibility associated with the nearby instability is already encoded in the defect response of the material.

More broadly, our work establishes mobile magnetic adatoms as sensors for imaging weak lattice distortions with atomic resolution. While spin excitations may provide an alternative read-out \cite{Amini2025}, YSR-state spectroscopy benefits from high energy resolution and a localized orbital character, providing access to hidden structural reconstructions that are typically inaccessible through topographic imaging or conventional tunneling spectroscopy. Our results therefore open a route towards studying incipient symmetry-breaking phenomena and the interplay between defects, lattice distortions, superconductivity, and charge order in quantum materials.

\section*{Acknowledgments}
We thank Carolina A. Marques, Dirk Morr, Jens Paaske, Antonio Sanna and Peter Wahl for discussions. We gratefully acknowledge financial support by the Deutsche Forschungsgemeinschaft (DFG, German Research Foundation) through Projects No. 277101999 (CRC 183, Project No. C03) and No. FR2726/10-1.

\section*{Data Availability Statement}
The data that support the findings of this article will be publicly available at Zenodo.

\bibliography{bibliography2}

\clearpage
\onecolumngrid

\beginsupplement
\section*{{Supplemental Material}}

\maketitle 

\section{Methods} \label{sec:methods}
All measurements were recorded using a low-temperature scanning tunneling microscope (SPECS JT-STM) operating at $T =$ \SI{1.1}{K}. Tips were prepared by chemically etching a tungsten wire and pre-characterized on bulk single crystals of Au(111) and V(110), which were cleaned using standard sputter and annealing cycles. Au-coated tips were obtained by repeated indentations into the Au surface. To enhance the energy resolution of superconducting gap spectra and realize atomic-scale Josephson spectroscopy, tips were coated with V by indentations into the V(110) surface. This procedure was repeated until the spectra on the substrate show a maximal double-gap size of $\sim$ \SI{1.50}{meV}. We note that the precise tip gap value may vary depending on the exact tip, but is unchanged within a given data set.

\NbS\ crystals were purchased from HQ Graphene and cleaved \textit{in-situ} under ultra-high vacuum conditions ($p$ $\sim$ \SI{e-10}{mbar}). We note that the cleaved surface tends to collect larger adsorbates and that the exposure time before transfer into the STM should be minimized to ensure large-scale pristine areas. Fe atoms were cold-deposited by e-beam evaporation, directly into the STM with an estimated sample temperature below \SI{12}{K}. Individual Fe atoms were laterally manipulated across the surface by approaching a V-coated tip on top of the atom to a setpoint of $\sim$ 2 to \SI{4}{nA} at \Vbias\ = \SI{5}{\mV}, moving laterally to the desired position and retracting the tip. 

We measure \dIdV\ signals using standard lock-in detection at a modulation frequency of \SI{811}{Hz} or \SI{919}{Hz} and a rms modulation amplitude of \SI{15}{\micro V} for spectra and \SI{0.5}{mV} for \dIdV\ maps, unless specified otherwise. Differential conductance maps of the YSR states were recorded in constant-contour mode: a height trace of the area of interest is recorded in constant-current mode at a given setpoint, and retraced while a different bias voltage is applied, and the current and \dIdV\ signal are recorded.

\section{Crystal and defect characterization}
Bulk NbS$_2$ appears in two polytypes, \NbS\ and $3R$-NbS$_2$, and only develops superconductivity in the $2H$-polytype \cite{Witteveen_2021}. Given the known challenge of growing phase-pure NbS$_2$ crystals \cite{Fisher1980}, we tested several crystals throughout this study and selected crystals with a well-developed superconducting gap. Each of these crystals also showed a similar density of intrinsic defects, as detailed below. Within the collection of studied superconducting \NbS\ crystals, we still found a varying sample quality, depending on the cleave and positioning of the tip on the crystal. For instance, in some cases we found step edges or sub-surface ripples in large-scale topographic images, which can induce non-local strain in the crystal and may influence the formation of a CDW phase \cite{Soumyanarayanan2013, Gao2018, Zhang2025}. Therefore, we selected crystals with flat, homogeneous surfaces and areas without step edges for this study. 

\begin{figure}[h]\includegraphics{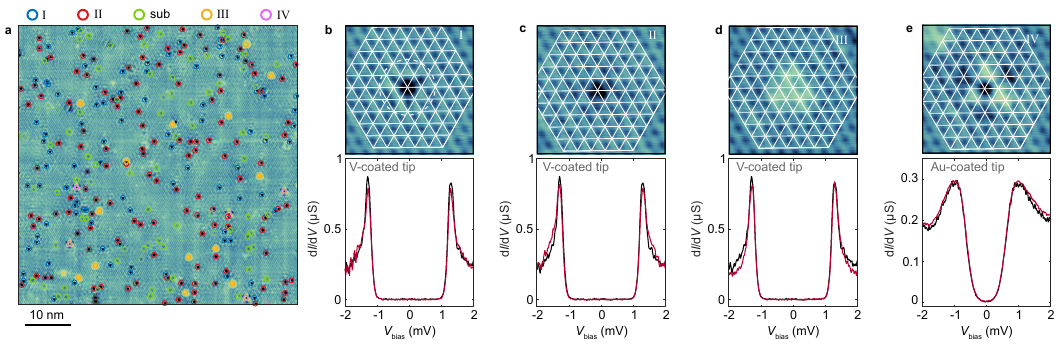}
	\caption{Defect statistics, binding site and spectroscopy. a) Large-scale topographic image of \NbS\ with marked defects (\It\ = \SI{1}{\nA}, \Vbias\ = \SI{100}{\mV}). The surface density of defects in this region is $n_{\Romannum{1}}$ = \SI{0.0361}{nm^{-2}}, $n_{\Romannum{2}}$ = \SI{0.0244}{nm^{-2}}, $n_{\Romannum{3}}$ = \SI{0.0039}{nm^{-2}} and $n_{\Romannum{4}}$ = \SI{0.0014}{nm^{-2}}. Additionally, fainter sub-surface defects are labeled, whenever visible, with $n_{\text{sub}}$ $\geq$ \SI{0.0214}{nm^{-2}}. b-e) Topographic images from Fig.\,\ref{fig:Fig1} of each defect type with white lines highlighting the atomic lattice, where each vertex corresponds to the position of a S atom (\It\ = \SI{1}{\nA}, \Vbias\ = \SI{100}{\mV}). Spectroscopic measurement on each defect type (red), compared to a substrate spectrum (black), are shown below each image. Spectra were stabilized at \It\ = \SI{1}{\nA}, \Vbias\ = \SI{5}{\mV} and recorded with a superconducting V-coated tip or a non-superconducting Au-coated tip ($V_{\text{mod}}$ = \SI{50}{\micro V}).} 
	\label{fig:S1_Defects}
\end{figure}

In Fig.\,\ref{fig:S1_Defects}a, a large topographic image of a pristine \NbS\ surface is shown with each defect type described in the main manuscript marked with a colored circle. Based on this image, we find that the surface density of defects amounts to $n_{\Romannum{1}}$ = \SI{0.0361}{nm^{-2}}, $n_{\Romannum{2}}$ = \SI{0.0244}{nm^{-2}}, $n_{\Romannum{3}}$ = \SI{0.0039}{nm^{-2}} and $n_{\Romannum{4}}$ = \SI{0.0014}{nm^{-2}}. Additionally, we mark the fainter triangularly-shaped depressions (green circles) and find $n_{\text{sub}}$ = \SI{0.0214}{nm^{-2}}. We note that this is a lower boundary, since they are only clearly identifiable in regions where other defects types are not present. Given their similar abundance to defect types \Romannum{1} and \Romannum{2}, we assign the fainter defects as sub-surface S vacancies. Based on the positions of all first-layer defects, the in-plane mean free path can be estimated by measuring the average travel distance before a defect site is reached. For an estimated scattering cross-section of $r = 2a$ (see dashed circle in Fig.\,\ref{fig:S1_Defects}b), we find that $l =$ \SI{8.1}{nm}. Notably, this is smaller than the superconducting coherence length, but larger than the Fermi wavelength \cite{Guillamon2008a, Heil2017}. 

The atomic resolution in Fig.\,\ref{fig:S1_Defects} can be used to superimpose a triangular grid of lines where each vertex corresponds to the bright contrast of a top-layer S atom. The grid reveals that defect types \Romannum{1}, \Romannum{2} and \Romannum{4} are centered around the position of a S atom. Given their abundance, defect types \Romannum{1} and \Romannum{2} are likely S vacancies, while type \Romannum{4} can be assigned to an substitutional or antisite defect. Defect type \Romannum{3} is less localized and centered around a hollow site, suggesting a defect in the Nb layer or an interstitial defect. Spectroscopic measurements of the superconducting gap on top of each of the defects (red lines in Fig.\,\ref{fig:S1_Defects}b-d) all show a full gap without additional in-gap states, demonstrating that there is no detectable magnetic moment on any of the defect sites. The rarest defect, type \Romannum{4}, was only probed with a non-superconducting tip.

\subsection{Quasi-particle interference of \NbS\ } \label{sec:dIdVanalysis}
\begin{figure}[h]
\includegraphics{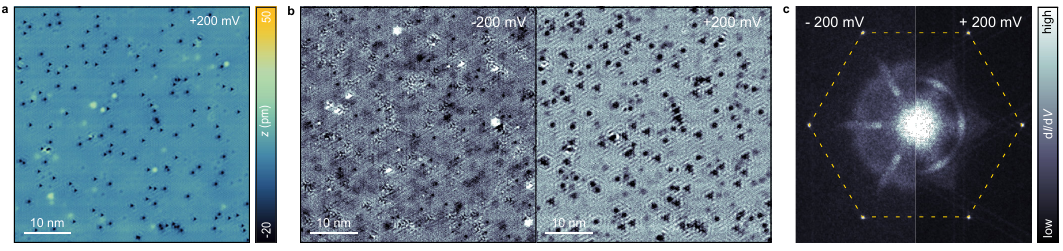}
	\caption{a) Topographic image of pristine \NbS, recorded at \Vbias\ = \SI{+200}{\mV},  \It\ = \SI{2}{nA}. b) Constant-current \dIdV\ maps recorded at \Vbias\ = \SI{-200}{\mV} and \SI{-200}{\mV}, and \It\ = \SI{2}{nA}, $V_{\text{mod}}$ = \SI{5}{mV}. c) Fourier-transformed image of \dIdV\ maps at \SI{-200}{\mV} (left half) and \SI{+200}{\mV} (right half). Fourier images were sixfold symmetrized to enhance the signal.} 
	\label{fig:SIFig_QPImap}
\end{figure}

The scattering of itinerant electrons with defects can lead to standing wave patterns, which can be imaged at the surface, typically referred to as quasiparticle interference (QPI) imaging. A key feature of QPI is that the periodicity of the pattern depends on the scattering vectors that connect different parts of the band structure, and therefore typically exhibits an energy-dependence that reflects the electronic band dispersion. To ascertain the origin of the observed wave patterns around intrinsic defects in \NbS, we measured two images of \SI{50}{nm} by \SI{50}{nm} regions on the pristine surface and recorded the \dIdV\ signal at \Vbias\ = \SI{+200}{\mV} and \SI{-200}{\mV}, as shown in Fig.\,\ref{fig:SIFig_QPImap}a,b. The Fourier-transformed images in Fig.\,\ref{fig:SIFig_QPImap}c show that the main scattering vectors project onto a ring-like feature with a diameter that decreases between \SI{-200}{\mV} and \SI{+200}{\mV}. This demonstrates that the observed pattern is dispersive and originates from electronic scattering.  

\section{YSR states of Fe atoms on \NoCaseChange{2\textit{H}-NbS$_2$}}
\subsection{Fe binding sites}
\begin{figure}[h]
\includegraphics{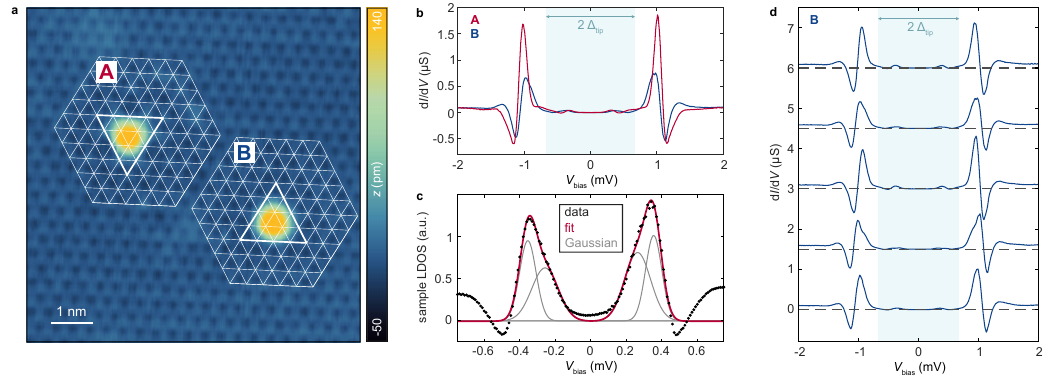}
	\caption{a) Topographic STM image of two Fe atoms on \NbS\ adsorbed in two different hollow sites (\Vbias\ = \SI{10}{\mV},  \It\ = \SI{50}{pA}). b) \dIdV\ spectrum of Fe$^{\text{A}}$ (red) and Fe$^{\text{B}}$ (blue) on \NbS, measured with a superconducting V tip (stabilized at \Vbias\ = \SI{5}{\mV},  \It\ = \SI{500}{pA}). The size of the superconducting gap of the tip is indicated with blue shading. c) Fe$^{\text{B}}$ spectrum (black) after numerical deconvolution, fitted with multiple Gaussian profiles (red) to identify two pairs of YSR resonances (gray). d) Collection of spectra on Fe$^{\text{B}}$ atoms positioned on various sites on the surface (artificially offset and stabilized at \Vbias\ = \SI{5}{\mV}, \It\ = \SI{500}{pA}).} 
	\label{fig:SIFig_Fe_bindingsites}
\end{figure}

The binding site of individual Fe atoms can be determined by analyzing their position with respect to the triangular lattice of S atoms, as shown in Fig.\,\ref{fig:SIFig_Fe_bindingsites}a. Analogously to Fe atoms on \NbSe\ \cite{Liebhaber2020}, we find that Fe atoms adsorb on two distinct hollow sites of the S lattice, depending on the relative position of the Nb atom below, which we label as Fe$^{\text{A}}$ and Fe$^{\text{B}}$ in Fig.\,\ref{fig:SIFig_Fe_bindingsites}a. The two Fe types exhibit a slight change in appearance, but more importantly, the change in crystal field environment results in a clear change in the YSR spectrum, as shown in Fig.\,\ref{fig:SIFig_Fe_bindingsites}b. The YSR resonances in the Fe$^{\text{B}}$ spectrum appear broader and with lower intensity compared to the Fe$^{\text{A}}$ type. The deconvolved spectrum of Fe$^{\text{B}}$ in Fig.\,\ref{fig:SIFig_Fe_bindingsites}c is well-described by two sets of Gaussian peaks. We find that the Fe$^{\text{B}}$ spectrum also shows significant variations across the surface, as exemplified for a collection of spectra recorded on six different positions in Fig.\,\ref{fig:SIFig_Fe_bindingsites}d. However, due to its sharp and well-defined $\gamma$ resonance, the Fe$^{\text{A}}$ type serves as a better atomic-scale sensor and was therefore selected to map the spatial variation in Fig.\,\ref{fig:Fig3}. 

\subsection{Deconvolution procedure}
\begin{figure}[h]
\includegraphics{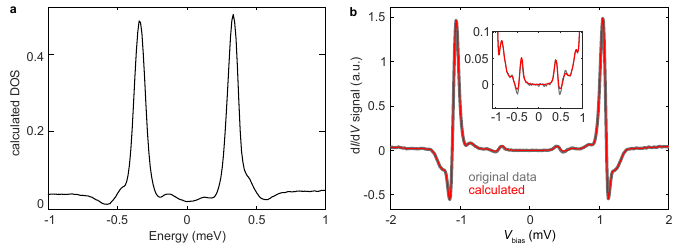}
	\caption{a) Sample DOS obtained after deconvolution of the spectrum in Fig.\,\ref{fig:Fig2}b. b) Original data (black) and calculated spectrum (red) by convoluting the spectrum in a) with the simulated tip gap spectrum. Inset: small energy window comparing the original and calculated thermal resonances. Parameters used for deconvolution: $\Delta_{\text{tip}} =$ \SI{0.74}{meV}, $\Gamma$ = \SI{0.008}{meV}, $T =$\SI{1.1}{K}.} 
	\label{fig:SIFig_Deconv}
\end{figure}

The detection of YSR states with a superconducting tip leads to an enhanced energy resolution by convolution of the sharp coherence peaks of the superconducting density of states of the tip with the in-gap resonance. The overlap of two sharp resonances can lead to negative differential conductance (NDC) in the spectrum, depending on the intensity of the YSR state. This feature can strongly influence the line shape of the in-gap resonance and may affect the appearance of other YSR states that are close in energy. A deconvolution procedure is therefore required to extract the local density of states of the sample and the line shape of each resonance, as shown in Fig.\,\ref{fig:SIFig_Deconv}a.  

The density of states of the tip is described by a standard Bardeen-Cooper-Schrieffer energy gap function, broadened by the Fermi-Dirac distribution and Dynes broadening \cite{Dynes1978}. Therefore, the deconvolution parameters are the tip gap size $\Delta_{\text{tip}}$, the Dynes parameter $\Gamma$ and the measurement temperature $T$. The validity of the deconvolution procedure is tested by comparing the original data with the convolution of the tip DOS and the calculated sample DOS. The parameters $\Delta_{\text{tip}}$ and $\Gamma$ are optimized to minimize the difference between the experimental and calculated data, as shown in Fig.\,\ref{fig:SIFig_Deconv}b.

\subsection{YSR maps}
\begin{figure}[h]	
\includegraphics{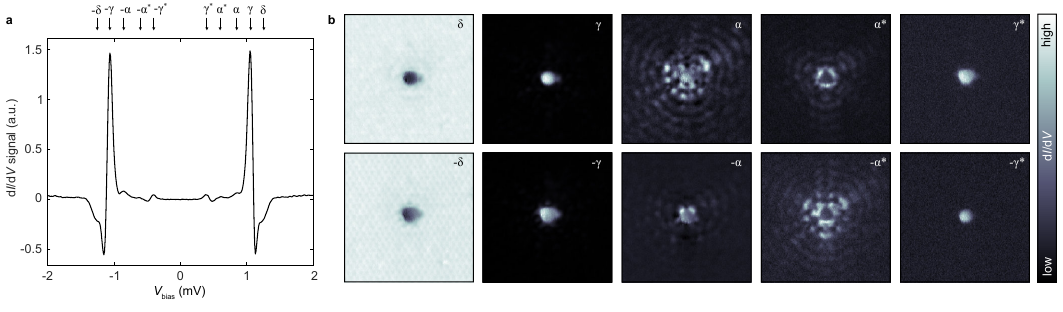}
	\caption{a) \dIdV\ spectrum of Fe on \NbS\ where the mapped energy is labeled for the YSR states $\alpha$, $\gamma$ and $\delta$, and the thermal resonances of $\alpha$ and $\gamma$, labeled with an asterisk (stabilized at \Vbias\ = \SI{5}{\mV},  \It\ = \SI{2}{nA}). b) \dIdV\ maps at each indicated energy, recorded in constant-contour mode at the height profile presented in Fig.\,\ref{fig:Fig2}a (recorded at \Vbias\ = \SI{5}{\mV},  \It\ = \SI{200}{pA}, $V_{\text{mod}}$ = \SI{30}{\micro V}).} 
	\label{fig:SIFig_YSRmaps}
\end{figure}

The spatial extent of each YSR state can be visualized by mapping the \dIdV\ signal at specific energies, as labeled in Fig.\,\ref{fig:SIFig_YSRmaps}a. In Fig.\,\ref{fig:SIFig_YSRmaps}b, we present \dIdV\ maps of the YSR states $\alpha$, $\gamma$ and $\delta$ at both polarities, as well as the thermal resonances of the $\alpha$ and $\gamma$ resonance. As expected, the spatial pattern of the thermal resonances reflects that of the YSR state at opposite polarity. We note that the spatial distribution seen in the $\delta$ and $-\delta$ images is heavily influenced by the NDC signal of the $\gamma$ resonance. 

\subsection{Influence of the tip}
\begin{figure}[h]
\includegraphics{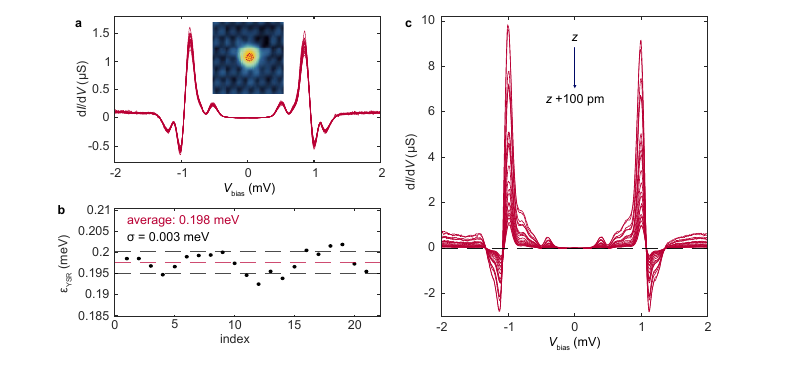}
	\caption{a) Collection of 21 \dIdV\ spectra on an individual Fe$^{\text{A}}$ atom, measured within a circle with a diameter of $\sim$ \SI{250}{pm}, as indicated in the inset (stabilized at \Vbias\ = \SI{5}{\mV},  \It\ = \SI{500}{pA}). b) Peak position of the $+\gamma$ YSR state extracted from the spectra in a), after deconvolution and Gaussian peak fitting. Dashed lines indicate the mean (red) and standard deviation (1$\sigma$) boundary lines (black). c) Collection of $z$-dependent spectra recorded centrally on an individual Fe atom. The tip was stabilized at \Vbias\ = \SI{5}{\mV}, \It\ = \SI{3}{nA} and retracted by \SI{5}{pm} steps between each spectrum in a total range of \SI{100}{pm}.}  
	\label{fig:SIFig_tipinfluence}
\end{figure}

The atomic-scale sensing approach presented in this work relies on detecting small variations in the YSR energy at each binding site. To test the influence of the tip position and the deconvolution procedure, we recorded a collection of spectra with the tip positioned at 21 different locations on the atom (see inset in Fig.\,\ref{fig:SIFig_tipinfluence}a). This mimics the potential variation of the exact tip placement during the measurement in Fig.\,\ref{fig:Fig3}. The resulting spectra in Figure \ref{fig:SIFig_tipinfluence}a show that the tip placement does not affect the YSR energy, but can lead to small variations in intensity. Additionally, we apply the aforementioned deconvolution and Gaussian fitting procedure to each of these spectra and extracted $\epsilon_{\text{YSR}}$ of the $\gamma$ state, as shown in Fig.\,\ref{fig:SIFig_tipinfluence}b. The distribution of extracted energies has a standard deviation of only \SI{3}{\micro V}, placing a lower bound on the statistically significant energy variations observed in Fig.\,\ref{fig:Fig3} and \ref{fig:Fig4}. We also measured the YSR spectrum of an individual Fe atom as a function of tip-sample distance in a range of \SI{1}{\angstrom} (Fig.\,\ref{fig:SIFig_tipinfluence}c). Naturally, the detected \dIdV\ signal increases with decreasing tip-sample separation, but the peak positions remain unaffected. 

\section{\NoCaseChange{\text{d}\textit{I}/\text{d}\textit{V}}  map analysis}
\begin{figure}[h]
\includegraphics{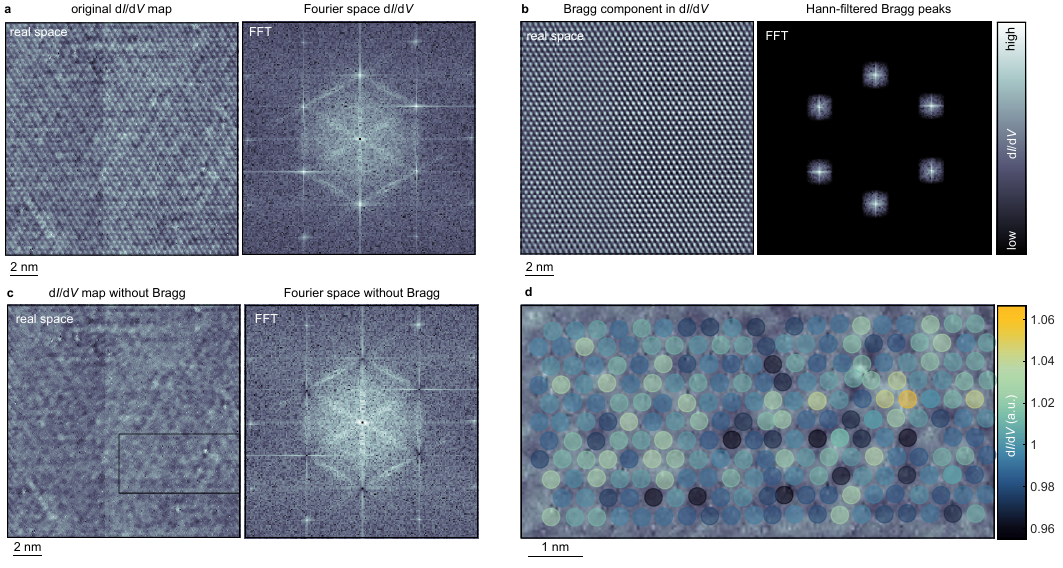}
	\caption{a) Original constant-current \dIdV\ image (left) and Fourier-transformed image (right). b) Real-space (left) and Fourier image (right) after applying a separable 2D Hann-window around each Bragg peak. c) Processed \dIdV\ map after subtracting the image in b) from the original image. d) Discretized \dIdV\ signal in the probed region from Fig.\,\ref{fig:Fig3} obtained by averaging the signal from the processed \dIdV\ image (background) at each binding site. This signal is normalized such that the color scale represents the percentage of observed variations. The image is recorded in constant-current mode at \Vbias\ = \SI{5}{\mV},  \It\ = \SI{1}{nA}, and processed with a Gaussian smooth filter to remove occasional spike artifacts.} 
	\label{fig:SIFig_dIdVmap}
\end{figure}

\begin{figure}[h]
\includegraphics{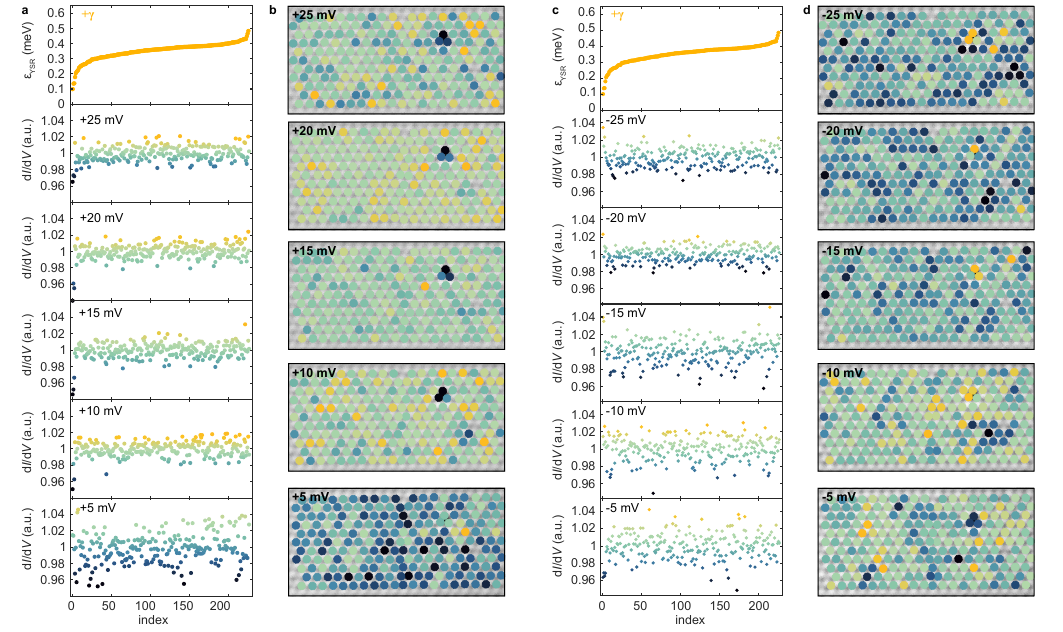}
	\caption{a) Sorted scatter plot of the YSR state energy $+\gamma$ with the corresponding average \dIdV\ signal at each binding site, extracted from \dIdV\ images recorded at \Vbias\ = \SI{25}{\mV}, \SI{20}{\mV}, \SI{15}{\mV}, \SI{10}{\mV}, \SI{5}{\mV}. b) Corresponding discretized \dIdV\ images at each energy. c) Sorted scatter plot of the YSR state energy $+\gamma$ with the corresponding average \dIdV\ signal at each binding site, extracted from \dIdV\ images recorded at \Vbias\ = \SI{-25}{\mV}, \SI{-20}{\mV}, \SI{-15}{\mV}, \SI{-10}{\mV}, \SI{-5}{\mV}. d) Corresponding discretized \dIdV\ images at each energy. Images are recorded in constant-current mode at \It\ = \SI{1}{nA} and $V_{\text{mod}}$ = \SI{1}{mV} was used for $|V_{\text{bias}}|$ $\geq$ \SI{20}{mV}.}
	\label{fig:SIFig_dIdVmap_disc}
\end{figure}

In order to compare the variation in $\epsilon_\gamma$ with the local variations of the density of states, we extract the \dIdV\ variations at each binding site. The \dIdV\ maps were recorded in constant-current mode in a \SI{16}{nm} by \SI{16}{nm} region that includes the investigated region in Fig.\,\ref{fig:Fig3}. The strongest signal, as seen in Fig.\,\ref{fig:SIFig_dIdVmap}a, originates from a topographic contribution at the atomic lattice periodicity. To isolate the variations in the LDOS, we apply Fourier filtering to remove the topographic information from the original image. Specifically, we apply a separable 2D-Hann window with a full-width half-maximum of 10 pixels around each of the Bragg peaks in the FFT image, which corresponds to a periodic lattice in real space (Fig.\,\ref{fig:SIFig_dIdVmap}b). This signal is subtracted from the FFT image to visualize the underlying LDOS variations as shown in Fig.\,\ref{fig:SIFig_dIdVmap}c. Lastly, we discretize the \dIdV\ signal by averaging the Fourier-filtered image within a circle with a \SI{150}{pm} radius at each binding site. The resulting set of points are color-coded and superimposed on the filtered \dIdV\ image in Fig.\,\ref{fig:SIFig_dIdVmap}d. This analysis was applied to ten \dIdV\ images, recorded at energies between \Vbias\ = \SI{25}{mV} and \SI{-25}{mV}, and compared to the sorted $\epsilon_{\gamma}$ in Fig. \ref{fig:SIFig_dIdVmap_disc}. Besides the binding sites directly at the defect site (first three indices), there is no observed correlation with the LDOS variation at any of the recorded energies.

\end{document}